\documentclass{sig-alternate-05-2015}

 \usepackage{subfigure}
  \usepackage{booktabs} 
  \usepackage{amssymb}
  \usepackage{graphicx}
  \usepackage{amsmath}
  \usepackage{mathrsfs}
  \usepackage{multirow}
  \usepackage{algorithm}
  \usepackage{algorithmic}
  \usepackage{url}
    \newtheorem{myDef}{Definition}
  
  \usepackage{enumerate}
\makeatletter
\def\@copyrightspace{\relax}
\makeatother
\begin{document}

\title{Heterogeneous Replica for Query on Cassandra}%

\numberofauthors{1} 
%

\author{
%
%
\alignauthor
Jialin Qiao, Xiangdong Huang, Lei Rui, Jianmin Wang\\
       \affaddr{Tsinghua University}\\
       \affaddr{Beijing, China}\\
       \email{{qjl16,ruil14}@mails.tsinghua.edu.cn,{huangxdong,jimwang}@tsinghua.edu.cn}
}

\maketitle
\begin{abstract}
Cassandra is a popular structured storage system with high-performance, scalability and high availability, and is usually used to store data that has some sortable attributes. When deploying and configuring Cassandra, it is important to   design a suitable schema of column families for accelerating the target queries. However, one schema is only suitable for a part of queries, and leaves other queries with high latency. 

In this paper, we propose a new  replica mechanism, called heterogeneous replica, to reduce the query latency greatly while ensuring high write throughput and data recovery. With this replica mechanism, different replica has the same dataset while having different serialization on disk. By implementing the heterogeneous replica mechanism on Cassandra, we show that the read performance of Cassandra can be improved by two orders of magnitude with TPC-H data set. 
\end{abstract}

\printccsdesc


\section{Introduction} 
\label{section1}
In recent years, with the development of the big data technology, big data applications are becoming more and more popular. For example, in the meteorological applications\cite{gui2012using}, a meteorological station will collect some metrics such as the temperature, humidity and wind speed at different dimensions such as latitude, longitude, altitude and timestamp for weather forecast \cite{xiang2014storage}. These dimensions are usually sortable and used to perform query filtering.


When using Cassandra to handle the above data in an application, developers have to collect the potential query patterns on the data and then design the database schema according to query patterns. For example, in Cassandra,  the partition keys \cite{clusteringkey} in a column family are used to partition data for load balance and can only be equality queried on. The clustering keys \cite{clusteringkey} support range filters or equality filters. Choosing different columns as partition and clustering keys will have different query performance. Therefore, users have to know all the potential query patterns before they design the database schema of Cassandra. 





In Cassandra,  the order of columns of clustering keys impacts query performance heavily. For example,  if the columns ({\it altitude, time})  are clustering keys, then queries such as ``{\it Find the humidity at some locations where altitude=100Pa and  between time [2018.05.01, 2018.05.02]}" have low latencies, because data is sorted by the {\it time} column on disk. If the columns ({\it time, altitude}) are clustering keys, then the latencies of the former queries increase but  queries such as ``{\it Find the humidity at some locations where altitude in [100Pa, 1000Pa] and time= 2018.05.01}  will have low latencies. This shows that organizing the column family structure efficiently accommodate one query, but may negatively impact the performance of many other queries \cite{bian2017wide}. Therefore, how to design the database schema according to the query patterns for optimizing the data serialization on disk is challenging.

In this paper, we rethink the ability of current replica mechanism in NoSQL systems \cite{han2011survey} and propose a new replica mechanism, called heterogeneous replica, to improve the query performance of the system. Currently, replica is used for data recovery and load balance of query. Traditional replica mechanism requires different replicas have the same serialization bytes on disk. Our heterogeneous replica only requires different replicas to have the same dataset while the serialization bytes on disk can be different. Different replica accelerates different query patterns. Meanwhile, the ability of data recovery and load balance are reserved because they have the same dataset. 

In Cassandra, the serialization of data is determined by the clustering keys in column family. By modeling the query cost on different schema of column families, we propose heterogeneous replica construction algorithm(HRCA) to construct the optimal heterogeneous replicas to minimize the average query latency for known query patterns. 

In this work, we adopt a general approach that separates architectural concerns of writing process and availability from heterogeneous replica. A shim layer heterogeneous replica (HR) engine was implemented on top of Cassandra \cite{lakshman2010cassandra} to verify the efficiency of our method. This enables the HR engine to control the exact heterogeneous replica structure on disk regardless of the write process of the underlying store.

The contributions of this paper are as follows:

\begin{itemize}
\item We propose a new replica mechanism called heterogeneous replica. The new mechanism gives replica the ability to accelerate queries while providing the original data recovery ability.
\item We propose a cost model for query on SSTable and formalize the heterogeneous replica construction problem to achieve the best query performance of given workload.
\item We propose HRCA algorithm to efficiently construct the optimal disk structures of heterogeneous replicas.
\item We implement a HR engine on top of Cassandra to verify the effect of our methods. Experiments show that we can achieve great performance improvement: the average query latency can be reduced by $1 \sim 2$ orders of magnitude.
\end{itemize}

The following of this paper is organized as follows. We first introduce the intuition in Section 2. The problem definition and the solution are introduced in Section 3. In Section 4 we show the architecture of our HR engine on Cassandra. The experimental evaluation is reported in Section 5. Finally, we discuss the related studies in Section 6 and conclude the paper in Section 7.

\section{Rethink REPLICA}
\label{section2}
Modern distributed storage systems usually use replica mechanism for data security, such as GFS \cite{ghemawat2003google, borthakur2008hdfs}, Cassandra \cite{lakshman2010cassandra}, Hbase \cite{wiki2012hbase}, MongoDB \cite{chodorow2013mongodb} and Dynamo \cite{decandia2007dynamo}. Replica mechanism is to copy data several times and store replicas on different nodes in distributed storage systems. When the data in one node is lost because of the disk crash, we can still recover the data through the copy of data on other nodes.

The side benefit of replica mechanism is that to support query parallelism and load balance. Queries on the same data can be evenly routed to different nodes which store a copy of the data. 

\begin{figure}[t]
    \centering
    \includegraphics[width=0.45\textwidth]{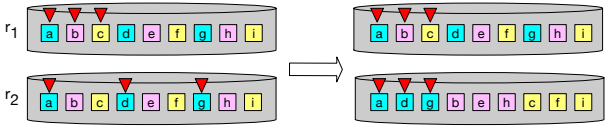}
    \caption{Heterogeneous Replica}
    \label{fig_hr}
    \end{figure}

Using traditional replica mechanism, all the replicas of a dataset have the same serialization bytes on disk, which have the same query performance for a query. The left part of Figure \ref{fig_hr} shows this case. Both replica $r_1$ and $r_2$ serialize the dataset $a\sim i$ on disk in alphabet order.  Given a query and either $r_1$ or $r_2$ serves for it,  the query latency is the same without considering load balance. 

The drawback is, a kind of serialization of replica on disk is only friendly for some queries. For example,   Given two queries in Figure \ref{fig_hr}, $q_1$ select data that is less than ``$d$" and $q_2$ selects  the ``blue" data.   the latency of $q_1$ is less than $q_2$ no matter which replica serves for the two queries.  That is, using traditional replica mechanism, no matter how we optimize the serialization strategy of the dataset on disk, one data serialization that accommodates some queries  (e.g., $q_1$) may negatively impact other queries (e.g., $q_2$). This ``{\it one size fits all}" pattern does not take full advantage of replica.

Our {\bf heterogeneous replica} mechanism only requires different replicas to have the same dataset while the serialization bytes on disk can be different. By this means, different replicas can handle different queries so that we can make full use of replica. The right part of Figure \ref{fig_hr} shows this case. $r_1$ serializes data in alphabet order and $r_2$ by color. Then both $q_1$ and $q_2$ have the minimal latency. At the same time, the data recovery ability is reserved because different replicas have the same dataset.

\section{Heterogeneous Replica}

\subsection{Modeling In Cassandra}
Heterogeneous replica mechanism cares for the data serialization on disk. In this paper we focus on Cassandra, which uses sorted sequence table (SSTable \cite{grigorik2014sstable,chang2008bigtable, shetty2013building}) to manage data on disk. SSTable stores key-value data in sorted order by the key. In Cassandra, given a partition key, the keys which are sorted in SSTable are the combination of clustering keys. That is, we can control the data serialization in a SSTable by adjusting the clustering keys of the corresponding column family. 

Then our problem is transferred to how to organize the order of clustering keys in each replica. We define that the order of clustering keys of a column family in a replica as the structure of the replica on disk.

We use $P=\{k_1,k_2,...,k_n\}$ to represent the columns that are chosen to be the clustering keys in a column family. $|P|$ is the number of columns in $P$. Given a dataset $D=\{d\}$, we use $d.k_i$ to represent the value of record $d$ at the $k_i$ column. For each column $k_i$, the distribution function of $\{d.k_i\}$  is $F_{k_i}(x)$ and the probability density function is $f_{k_i}(x)$.  $A=\{ck_1,ck_2,...,ck_n\}$ is a permutation of clustering keys, i.e. $ \forall i \in [1,n], \exists j, k_i = ck_j$. In practice, $A$ is the structure of a replica.

We use $Q=\{q_1,q_2,...q_n\}$ to represent the known query workload. Each query pattern is composed of some range filters ($\{d | d.ck_i\in [s_i, e_i)\}$ ) and equality filters ($\{d | d.ck_i = v_i\}$) on clustering keys $A$. For example, $q_3=\{d | d.ck_1=4 \land d.ck_2\in[5, 8) \land d.ck_3\in [3, 8) \}$ . Besides, we assign the clustering keys that do not has any filter with a global range filter to ensure that every clustering key has a filter. To support these queries, we use ALLOW FILTERING \cite{allowfiltering} in Cassandra.

\begin{figure}[t]
    \centering
    \includegraphics[width=0.4\textwidth]{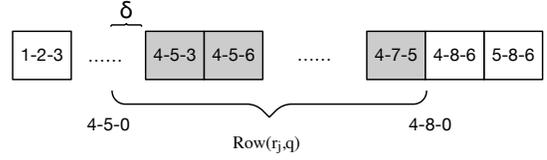}
    \caption{Rows need to be read from SSTable}
    \label{fig_estimate}
    \end{figure}

Given a query, the time cost mainly  depends on the size of data in SSTables to be loaded from disk in Cassandra.  For example, Figure \ref{fig_estimate} shows the data needs to be loaded of a SSTable when executing $q_3$, in which $ck_1-ck_2-ck_3$ is the combination of clustering keys. Cassandra needs to traverse from the lower bound (4-5-3) and terminate when meet the first key (4-8-6) that exceeds the end boundary. 

When estimating the size of these rows, we record the index of the first clustering key that has an range filter as $i$, ensuring all keys before the i-th key has an equality filter. Given a replica $r_j$ with the clustering key permutation $A$, the number of rows to be loaded from disk can be estimated as:

	\begin{equation}
	\label{equ_row}
	Row(r_j,q) = |P| * \prod^{i-1}_{p=1} f_{ck_p}(v_p) * (F_{ck_i}(e_i)-F_{ck_i}(s_i))
	\end{equation}

This estimation has a little larger $\delta$ compared to the real size of rows shown in Figure \ref{fig_estimate}. 

In the query process of Cassandra, the real time cost $Cost()$ is dominated by $Row()$ and the relation between $Row()$ and $Cost()$ depends on the actual environment of the system. We use $f()$ to represent:

	\begin{equation}
	\label{equ_cost}
	Cost(r_j,q) = f(Row(r_j,q))  
	\end{equation}


Given a specific structure of $N$ replicas $R=\{r_1,r_2,...,r_N\}$ on disk and a query $q$, The minimal time cost of a query is as following

	\begin{equation}
	Cost_{min}(q) = \{Cost(r_i,q) | \nexists j \in [1,N] , Cost(r_j,q)<Cost(r_i,q)\}
		\label{equ_cost_R}
	\end{equation}
	
Then, the average time cost of $Q$ is:

	\begin{equation}
	Cost(R,Q)=\frac{1}{|Q|}\sum_{i=1}^{|Q|}(Cost_{min}(q_i))
	\label{equ_cost_rq}
	\end{equation}

Finally, the HRC problem is defined as follows:

	\begin{myDef}
    	(Heterogeneous replica construction problem): Given a query workload $Q$, find optimal structure of heterogeneous replicas $R^*$ such that the average latency of $Q$ is minimized
	
	\begin{equation}
	\label{equ_goal}
	R^*=arg \min \limits_{R}  \{ Cost(R,Q) \}
	\end{equation}
	\end{myDef}

\subsection{Replica Construction}

In this section, we first analyze the hardness of the HRC problem, which motivates us to devise an HRCA algorithm based on simulated annealing.

The simple way to find $R^*$ is to enum all possible structures of replicas $R$ in Equation (\ref{equ_goal}). Given the replication factor $n$ and the number of clustering keys $m$,  $C^n_{m!+n-1}$ kinds of  replica layouts could be considered as possible results. It can be very large when $m$ or $n$ are large. 

Therefore, we propose an algorithm to find an approximation of the optimal heterogeneous replicas based on simulated annealing \cite{Kirkpatrick1983OptimizationBS}. Algorithm \ref{alg_replica_construction} shows the details.

\begin{algorithm}[t] 
	\caption{Heterogeneous Replica Constructing Algorithm} 
	\label{alg_replica_construction} 
	\begin{algorithmic}[1] 
		\REQUIRE ~~\\ 
		$Q$: Query workload\\
		$R_0=\{r_1,r_2,...r_N\}$:The initial structure of replicas \\
		\ENSURE ~~\\ 
		$R$: The optimized heterogeneous replicas
		\STATE $t :=t_0, R := R_0, C := Cost(R_0,Q)$
		\FOR{$k=1 \ to \ k_{max}$} \label{line_start}
			\STATE $R' := NewState(R)$
			\STATE $ C' := Cost(R',Q)$  \label{line_cost}
			\IF{$C' < C || e^{\frac{C - C'}{t}} > random(0,1)$}
				\STATE $R := R'$, $C := C'$
			\ENDIF
		\ENDFOR \label{line_end}
		\RETURN $R$
	\end{algorithmic}
\end{algorithm}

In HRCA, a specific structure of all heterogeneous replicas $R$ on disk corresponds to a \textbf{state}.  Users need to give a query workload $Q$ and arbitrary state as the initial state $R_0$. The main loop (line \ref{line_start}-\ref{line_end}) of Algorithm \ref{alg_replica_construction} is the searching process in simulated annealing.  A `good' state will always be accepted, while a `bad' state will be accepted probabilistically, which can avoid finding a locally optimal solution. The new state generation function $NewState(R)$ is generated by swapping two clustering keys of a replica $r_j$ in $R$.  

The algorithm is only be called once so the speed is not important compared to the effect. Besides, the algorithm generally converges in ten seconds in our experiments.

\section{Architecture}
\label{sec_imp}

\subsection{Separation of Concerns}

We implemented the heterogeneous replication(HR) engine on top of Cassandra. This architecture decouples the replica mechanism from the data management on disk. 

This enable a clean separation of concerns. The underlying Cassandra handles most aspects of data management, including the management of MemTable and SSTable process of LSM-Tree\cite{o1996log, rockdb2017cidr}. Hence we do not worry about the sorting and data serialization on disk as well as the compaction strategy of SSTables. The above HR-engine can concentrate on how to construct an optimal structures of heterogeneous replicas.

The underlying Cassandra handles most aspects of data management, including the management of MemTable and SSTable process of LSM-Tree. Hence we do not worry about the sorting and data serialization on each node as well as the compaction strategy of SSTables. 






\subsection{HR Engine}

\begin{figure}[t]
    \centering
    \includegraphics[width=0.45\textwidth]{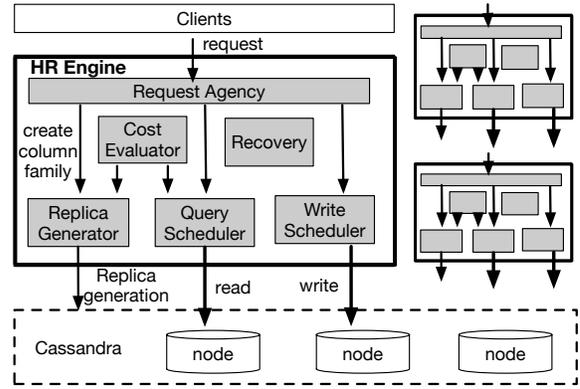}
    \caption{HRCA Engine}
    \label{fig_implement}
    \end{figure}

The architecture is shown in Figure \ref{fig_implement}. HR engine mainly has five models: request agency, cost evaluator, replica generator, request scheduler and recovery. The engine accepts the requests of clients and connect to the underlying database.

All requests from clients are routed by the \textbf{Request Agency}. Clients send requests to the \textbf{Request Agency} layer, which handles all communications with underlying data store and other clients. Clients are agnostic to the Cassandra.

For {\it CREATE COLUMN FAMILY} request, the \textbf{Replica Generator} module automatically generates the optimal replica structures and allocates to different nodes by our defined hash function that accept the replica id and partition key. 

When receiving a read request, the \textbf{Request Scheduler}  will route the request to the replica with lowest latency. The query cost in a replica will be calculated by the \textbf{Cost Evaluator}. The write request will be resolved by the \textbf{Write Scheduler} and be sent to all replicas to maintain the data consistency. The sort process will be handled by the LSM-Tree writing process of each node.

\textbf{Recovery} is responsible for data recovery when a node down. As the structures of replicas are different, the original recovery strategy does not apply to the heterogeneous replica. Therefore, we leverage the LSM-Tree write process to recover replica. 



\section{Evaluation}
In this section, we first model the cost function $f()$ in Equation (\ref{equ_cost}). Then we compare the following replica mechanism under different data size, replication factor, and the number of clustering keys on TPC-H and simulation dataset:

\begin{itemize}
\item \textbf{TR} The traditional replica mechanism with approximate optimal structure that an expert can give.
\item \textbf{HR} The heterogeneous replicas that HRCA generates.
\end{itemize}

We ran experiments on a Cassandra (version 3.11.0)  cluster with 6 nodes. Each node has 2 Intel Xeon E5-2697 CPUs which have 36 cores in total, 256GB memory and 7200 rpm HDD.

There are two kinds of datasets we used:

\textbf{TPC-H Dataset} Considering the column data type and supported queries, we use the table \textbf{orders} in TPC-H, which has 9 columns as the experiment target.  The scale factor (of TPC-H dataset) we used is $1 \sim 5$ which results in different data size ranging from 1.5 million to 7.5 million. In this column family, the clustering key we defined are {\it custkey}, {\it orderdate} and {\it clerk}.

Since the 22 queries in TPC-H are mainly for join operations which are not the optimization target of our query model, we give two examples based on the business scene:

\begin{itemize}
\item \textbf{Q1}: find the total price of the all customers that a specific clerk served at one day. SQL Example : {\it select totalprice from orders where orderdate = ? and clerk = ? and custkey $\geq$ 0};
\item \textbf{Q2}: find the total price that a customer consumed by a specific clerk's merchandising in some days.  The time range of {\it orderdate} are randomly generated. SQL Example : {\it select sum(totalprice) from orders where custkey = ? and clerk = ? and orderdate $\geq$ ? and orderdata < ?};
\end{itemize}

We generated 500 query instances by replacing the ``{\it ?}" in the SQL templates. 

{\bf Simulation dataset}: We generated simulation datasets, whose data size $|P|$ and clustering key size $|D|$ satisfy: (1) The value scope in each clustering key is $0 \sim \log_{|D|}|P|$; and (2) the data type of each clustering key is integer and is distributed randomly in the whole data space. The queries we used is randomly generated.

\begin{figure}[t]
	\centering 
	\subfigure[size of row value]{
		\label{fig_value_costmodel}
		\includegraphics[width=0.4\linewidth]{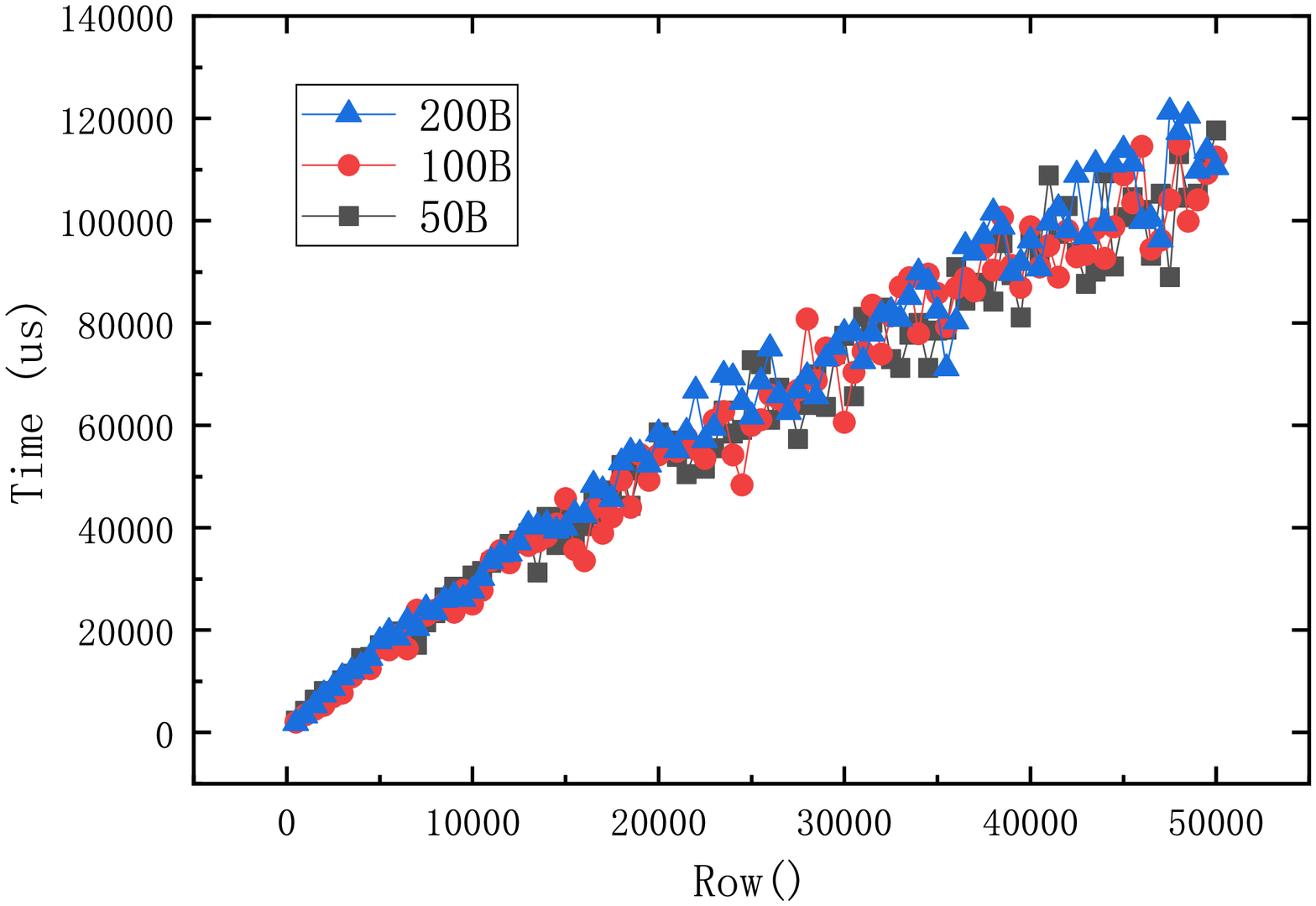}}
	\hspace{0.01\linewidth}
	\subfigure[number of clustering keys]{
		\label{fig_column_costmodel}
		\includegraphics[width=0.4\linewidth]{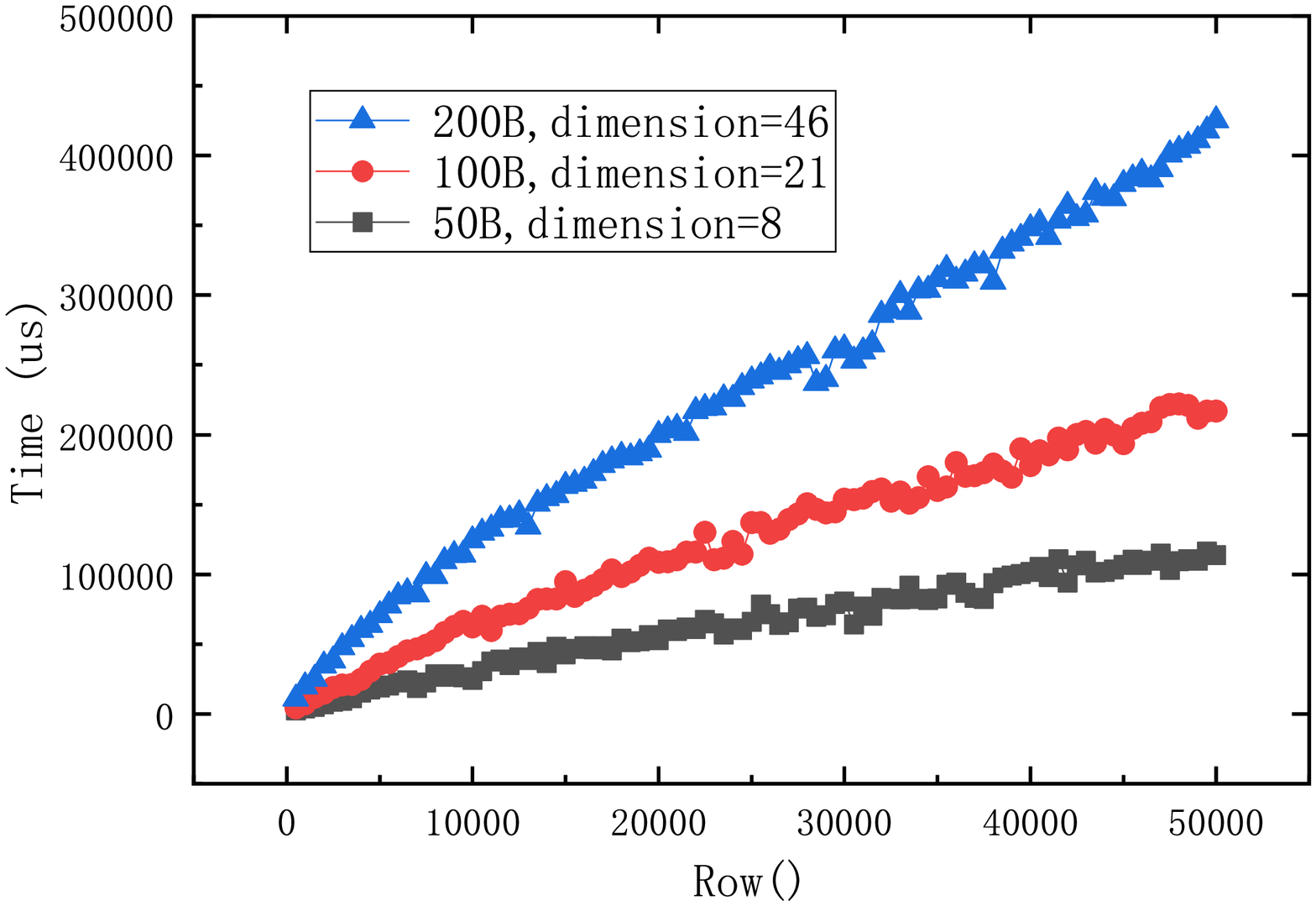}}
	\caption{$f()$ with different parameters}
	\label{fig:subfig2}
\end{figure}

\begin{figure*}[t]
	\centering 
	\subfigure[Latency with different data sizes on TPC-H dataset]{
		\label{fig_datasize_time}
		\includegraphics[width=0.3\linewidth]{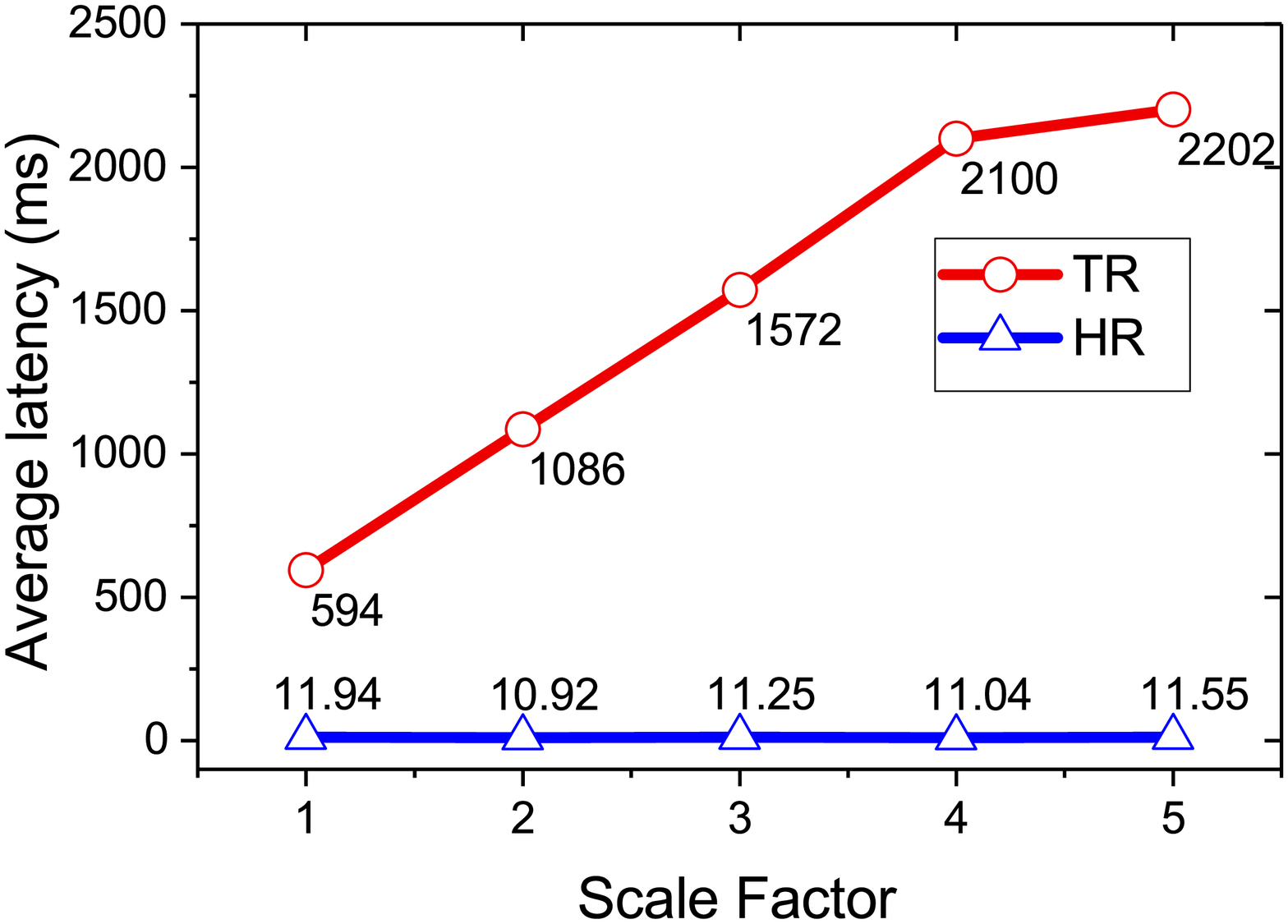}}
	\hspace{0.01\linewidth}
	\subfigure[Latency with different replication factors on simulation dataset]{
		\label{fig_rf_time}
		\includegraphics[width=0.3\linewidth]{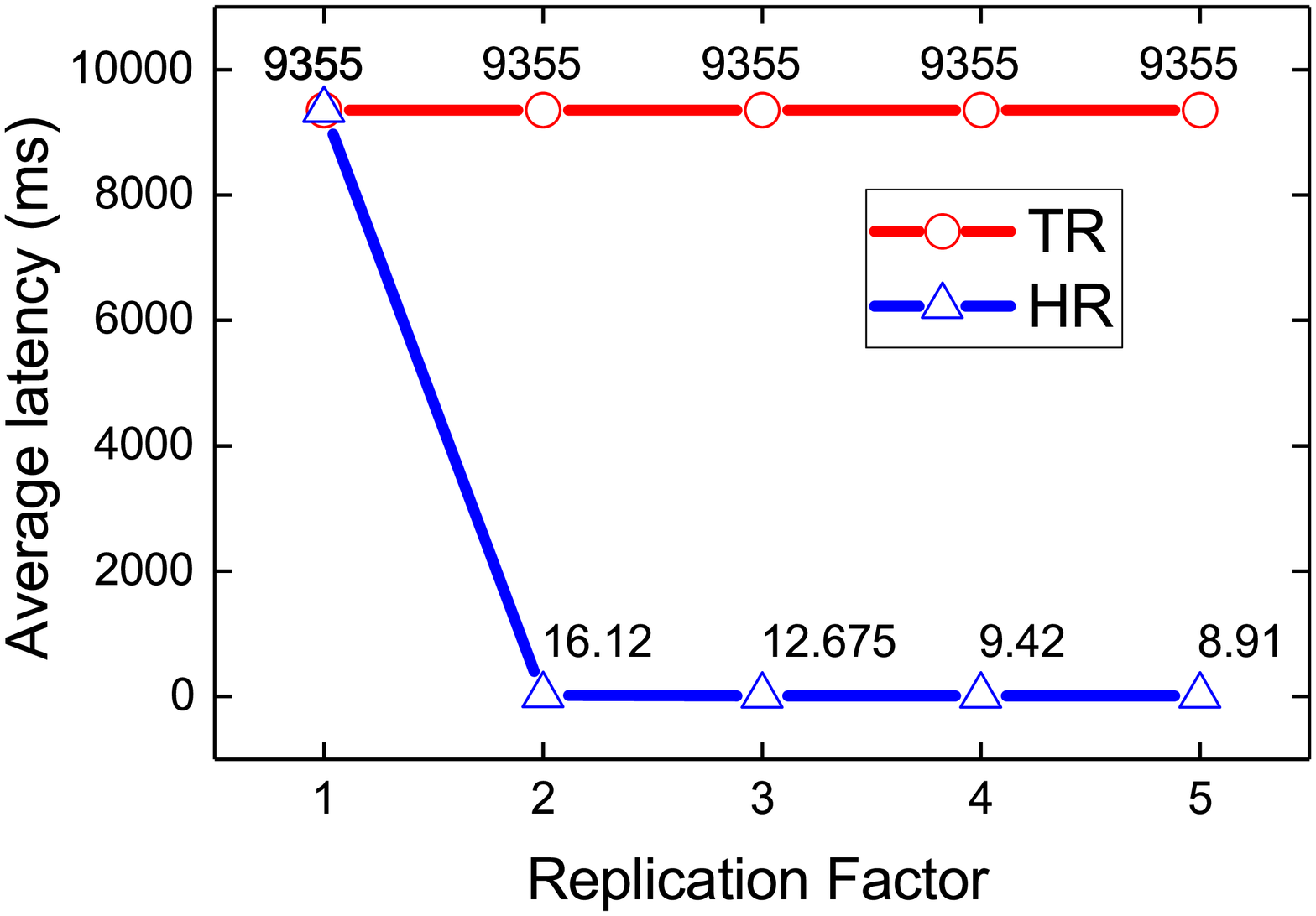}}
	\subfigure[Latency with different clustering key sizes on simulation dataset]{
		\label{fig_clustering key_time}
		\includegraphics[width=0.3\linewidth]{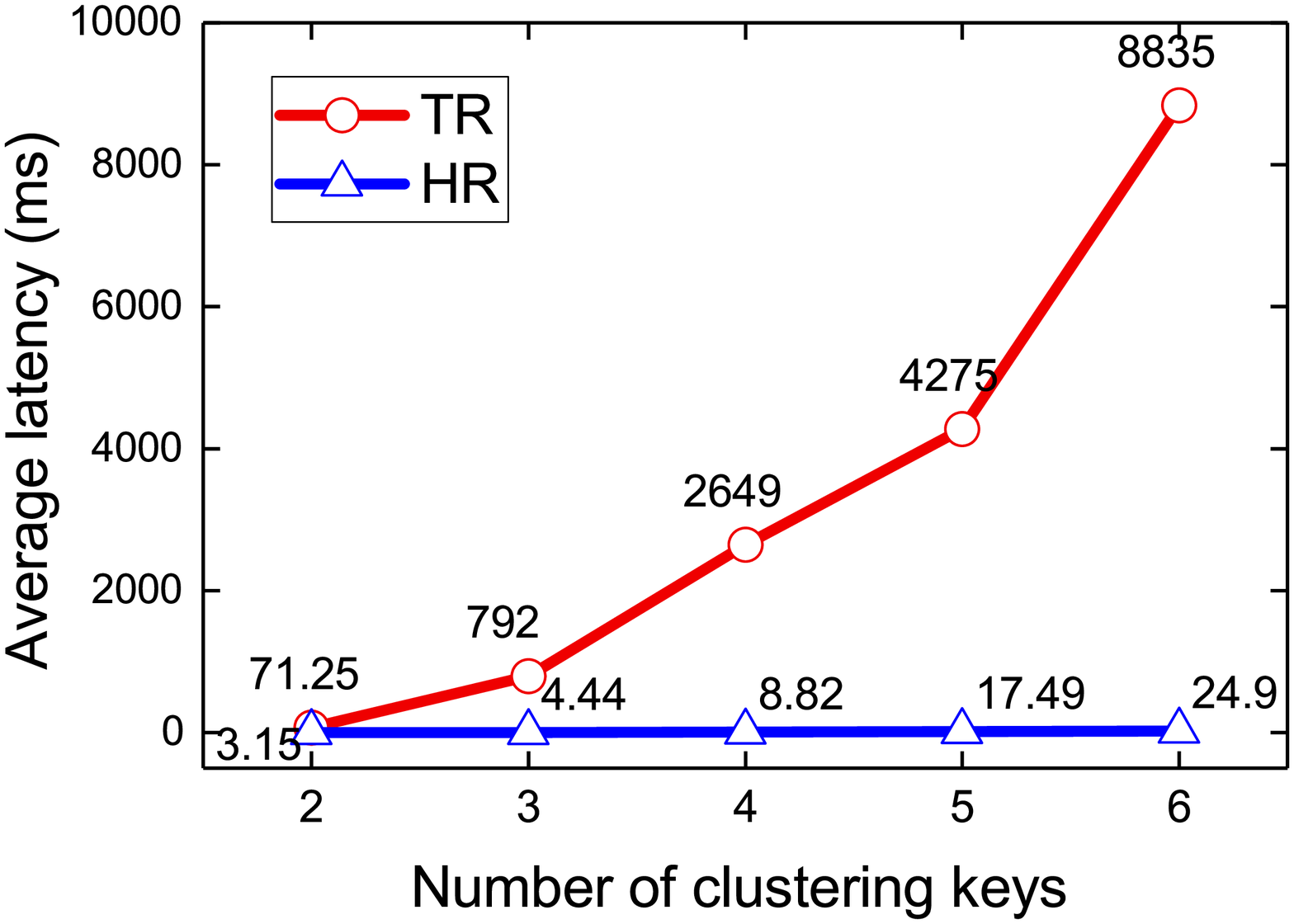}}
	\vfill
	\subfigure[Gain with different data sizes on TPC-H dataset]{
		\label{fig_datasize_improve}
		\includegraphics[width=0.3\linewidth]{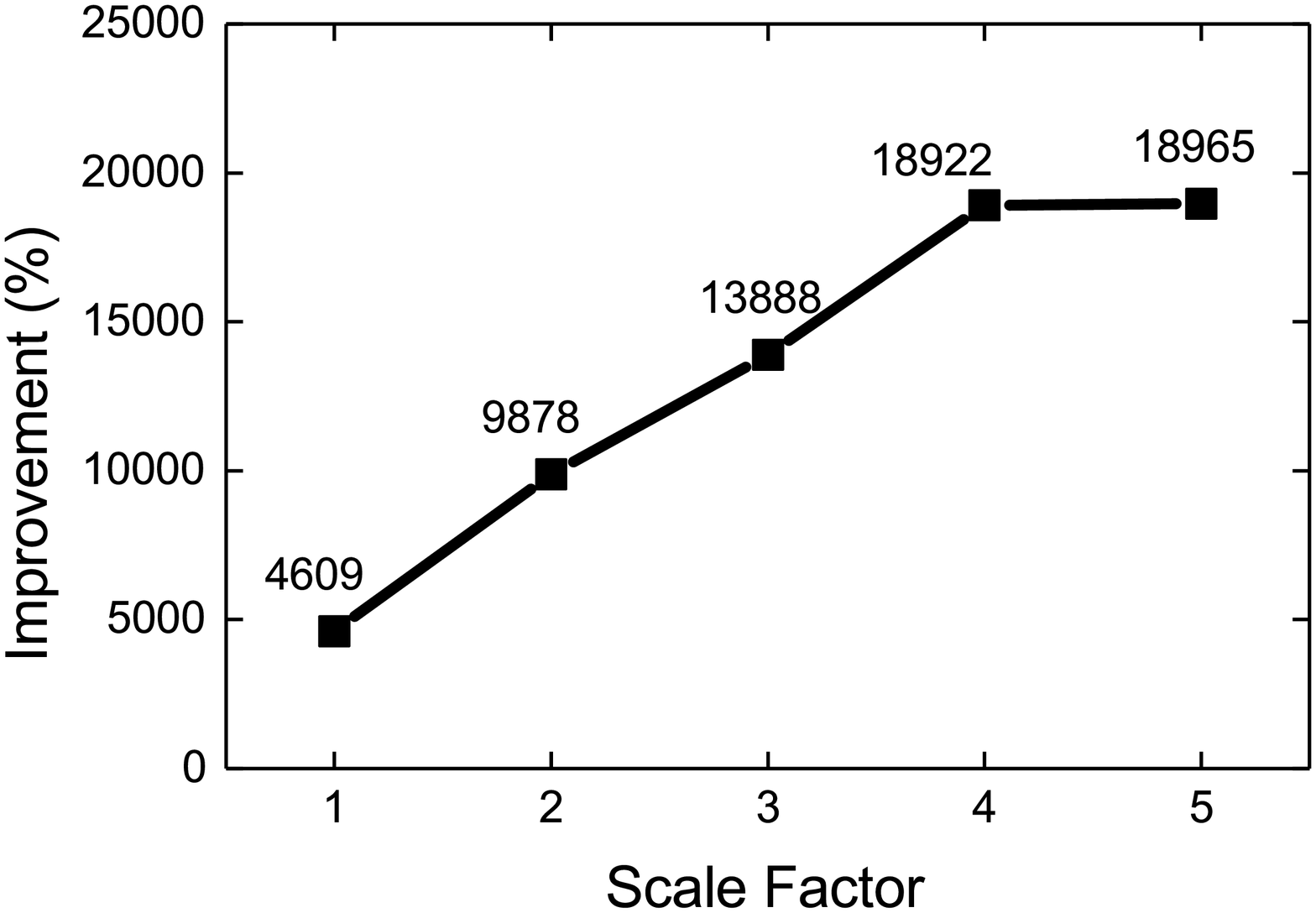}}
	\hspace{0.01\linewidth}
	\subfigure[Gain with different replication factors on simulation dataset]{
		\label{fig_rf_improve}
		\includegraphics[width=0.3\linewidth]{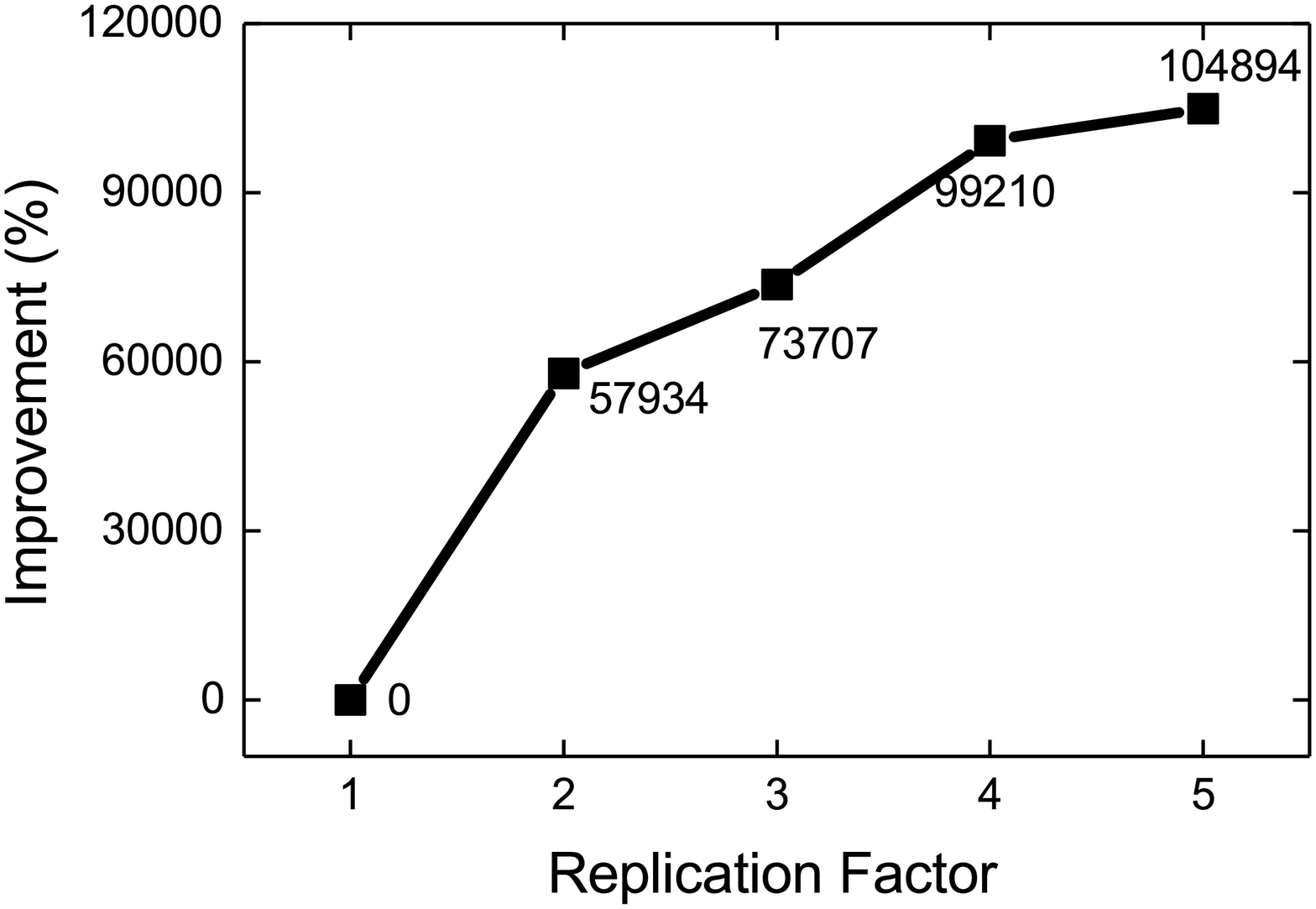}}
	\subfigure[Gain with different clustering key sizes on simulation dataset]{
		\label{fig_clustering key_improve}
		\includegraphics[width=0.3\linewidth]{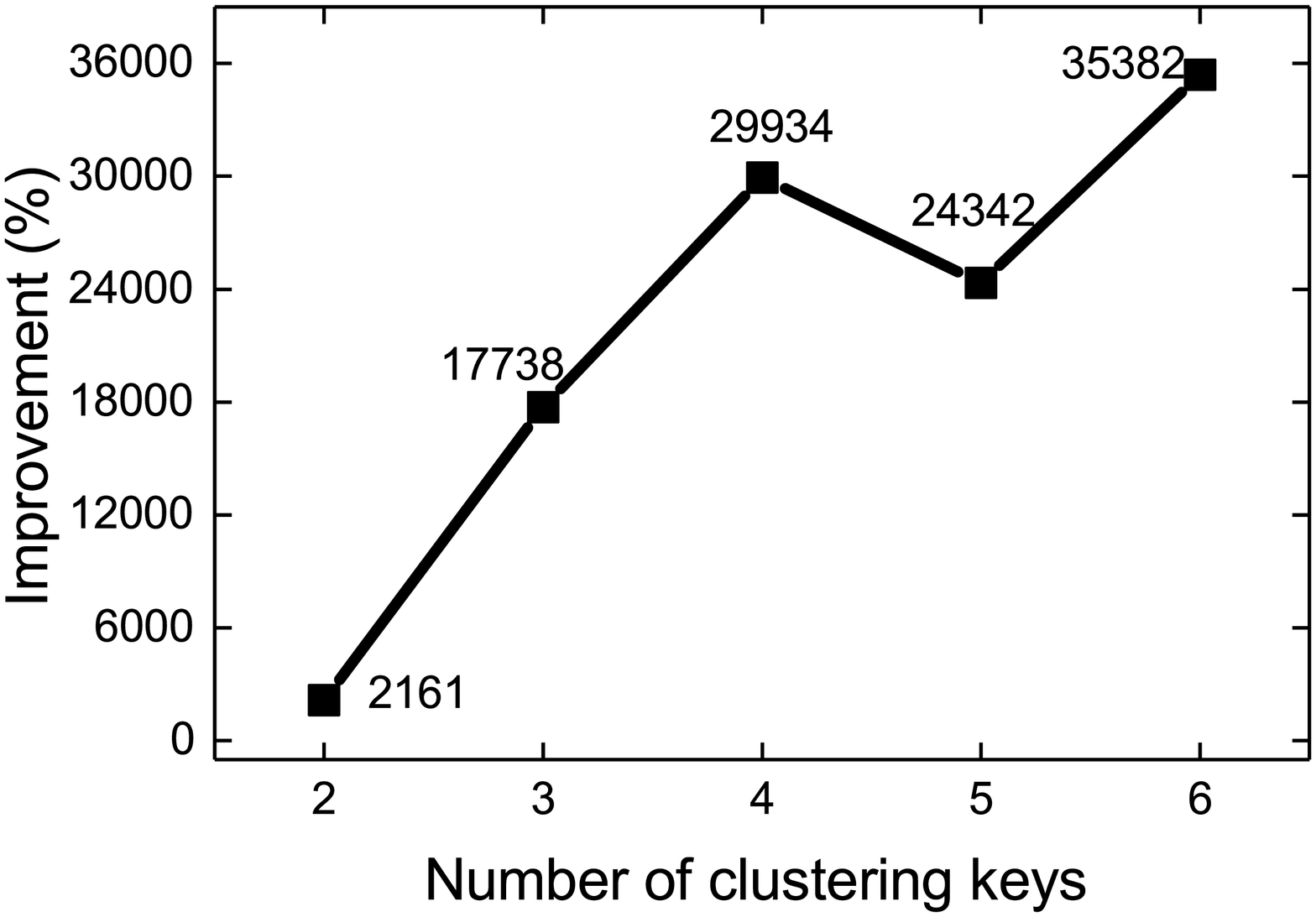}}
	\caption{Query latency and gain of HR with different parameters}
	\label{fig:subfig}
\end{figure*}

\subsection{Cost Modeling}

In Equation (\ref{equ_cost}), the function $f()$ depends on the hardware and the configurations of the system. In this experiment, we collect the query time costs under different $Row()$ in different system environments using the simulation dataset with queries.  The result is shown in Figure \ref{fig:subfig2}.

We first  evaluate whether the size of data item impacts Row(). In the experiment, we change the size of the data item from 50B to 200B by increasing the size of the metric column value. As shown in Figure \ref{fig_value_costmodel}, each line represents the cost with a specific size of a data item. The lines  show the roughly linear relationships between the size of candidate result set $Row()$ and the cost. Besides, the cost does not change significantly while the size of a data item increases three times(i.e., from 50 byte to 200 byte). Therefore, we do not need to model the cost function if only the  size of the metric column changes.

We then study how the number of clustering keys impacts the cost function. The number of clustering keys is also related to the size of data items, so we change the number to make the average item size be 50 bytes $\sim$ 200 bytes as the above experiment. As can be seen in Figure \ref{fig_column_costmodel}, the cost function is a linear function under different number of clustering keys. The slope of the cost function increases when the number of clustering keys is increasing. Therefore,  it needs to be re-modeled.

\subsection{Query Latency and Improvement}

We evaluate the average query latency of two replica mechanisms under different size of the TPC-H datasets.  The result is shown in Figure \ref{fig_datasize_time} and the relative performance improvement of HR over TR is in Figure \ref{fig_datasize_improve} using the equation $\frac{Cost(TR)-Cost(HR)}{Cost(HR)}$. The query cost of the TR mechanism increase gradually as the data size grows while the HR remains almost unchanged. The improvement from TR to HR shows the good performance of heterogeneous replicas. According to the experiment, HR can achieve two orders of magnitude performance improvement compared to the TR under 5 scale factor.

We then evaluate how replication factor(number of replicas) impacts the query latency on the simulation dataset. In this experiment, there are 10 million data items in the dataset. The query latency of two mechanisms with replication factor from $1 \sim 5$ are shown in Figure \ref{fig_rf_time}. The relative performance improvement of HR over TR is shown in Figure \ref{fig_rf_improve}. The average query latency stays constant in TR but decreases as the replication factor grows in HR. The latency is the same when the replication factor is 1 and the time cost of HR dramatically drops when the replication factor is greater than 1. Two replicas can greatly speed up the query.

We evaluate the impact of clustering key number on the simulation dataset with uniform queries by changing the number of clustering keys $|D|$ from $2 \sim 6$. The dataset has 10 million items in total and the replication factor is 3. The time costs of different replica mechanisms are shown in Figure \ref{fig_clustering key_time} and the relative performance improvement of HR over TR is shown in Figure \ref{fig_clustering key_improve}.  As can be seen, the improvement increases along with clustering key increasing except for five clustering keys. If there are only 2 or 3 clustering keys, the effect of 3 replicas is not fully utilized. For more clustering keys, the effect of HR is better.

\subsection{Write Throughput}

We measured the write throughput of TR and HR with 3 replicas on TPC-H dataset. We load 40, 80 and 120 million rows separately. Table \ref{tab:write} shows that heterogeneous replica maintain the same write speed as traditional replicas. Because we write data asynchronously into different replicas and the writing process of different replica use the traditional LSM-Tree write strategy. Therefore, heterogeneous replica does not have a negative impact on write speed.

\begin{table}[]
\caption{Time cost on loading data}
\label{tab:write}
\resizebox{0.5\textwidth}{!}{%
\begin{tabular}{|l|l|l|ll}
\cline{1-3}
Number of Rows(million) &  Time cost with TR(s) & Time cost with HR(s) &  &  \\ \cline{1-3}
40                      & 819                   & 823                   &  &  \\ \cline{1-3}
80                      & 1539                  & 1533                  &  &  \\ \cline{1-3}
120                     & 2259                  & 2257                  &  &  \\ \cline{1-3}
\end{tabular}%
}
\end{table}

\subsection{Data Recovery}

We measure the speed of data recovery when a node falls down. We remove the data on the node and call {\it nodetool repair} to launch the origin data recovery in Cassandra. We import 18 million rows of TPC-H data set, traditional data recovery takes approximately 4 minutes to recover data, our HR-engine takes 6 minutes to recover. Considering that the node failure occurs infrequently, compared to the tremendous reduction of query latency, take a little longer to recover data is acceptable.

\section{RELATED WORK}

There are many works \cite{huai2013understanding, jindal2013comparison} to optimize the data structure on disk. For example, in columnar store \cite{melnik2010dremel}, a column ordering strategy is proposed in \cite{bian2017wide}. The read speed is improved by adjusting the disk order of columns. They model the disk seek cost on column store but we model the data needs to be loaded into memory on SSTable. Besides, they do not consider replica.


Data duplication and replica mechanism \cite{sanders2001denormalization, lei2008line, saito2005optimistic, van2004chain}  have been widely used in data management systems. In which full replication \cite{kemme2000don, plattner2008extending} is a popular approach. For example,  primary-backup \cite{budhiraja1993primary}. The main goal of full replication is supporting the data availability. In \cite{rabl2017query}, data replication and allocation strategy are proposed. This work chooses optimal replication factors for each partition. Another work \cite{zhong2010dynamic} increases replica according to the frequency of the query access. In  another work \cite{bian2017wide}, a duplication strategy is proposed for columnar store systems.  The above works increase either the number of replicas or the amount of data in each replica. Some even assume that the storage is unlimited. We make adjustments based on the existing replica mechanism, and improves the query performance without introducing additional disk costs, which is different from existing works.

Data partitioning methods place different part of data on different nodes to achieve load balancing and thereby accelerate the query. We focus on the heterogeneous replica structure inside each partition. Therefore, data partitioning strategy are orthogonal to our work and can work with our replica mechanism together.

\section{Conclusions}
In this paper, we propose a new replica mechanism called heterogeneous replica. The new mechanism gives replicas the ability to significantly reduce the average latency of queries while keeping the properties of data recovery. The existing approaches on accelerating queries either optimize limited kinds of queries by adjusting the structure of data or duplicate some frequently accessed data. In contrast with them, we do not introduce additional disk cost by optimizing the existing replica serialization on disk. 

To find an approximate optimal structures of the heterogeneous replicas, we propose (1) a cost model for SSTable on Cassandra, (2) the formalized heterogeneous replica construction problem, (3) a solution to find the optimal structures of replicas, (4) the implementation of HR engine.  We believe that our replica mechanism can be also applied to other databases not only Cassandra that have replica. 

\bibliographystyle{abbrv}
\bibliography{heterogeneous_replica}

\end{document}